\documentclass[aps,prl,twocolumn,floatfix,preprintnumbers,showpacs]{revtex4}
\usepackage{graphicx}
\usepackage{amsmath}
\usepackage{amsfonts}
\usepackage{color}
\usepackage[dvips]{draftcopy}

\newcommand{\etal}{{\em et al.}}
\newcommand{\dn}[2]{d^{#1}{#2}\,}

\newcommand{\rmsubscript}[2]{{#1}_{\textrm{#2}}}

\newcommand{\Rside}{\rmsubscript{R}{side}}
\newcommand{\Rout}{\rmsubscript{R}{out}}
\newcommand{\Rlong}{\rmsubscript{R}{long}}

\begin{document}

\preprint{\rm UCRL-JRNL-215137-DRAFT}

\title{
      Exploring Lifetime Effects in Femtoscopy
}
\author{D.A.~Brown} 
\affiliation{Lawrence Livermore National Laboratory, Livermore California 94551}
\author{R. Soltz}
\affiliation{Lawrence Livermore National Laboratory, Livermore California 94551}
\author{J. Newby} 
\affiliation{Lawrence Livermore National Laboratory, Livermore California 94551}
\author{A. Kisiel}
\affiliation{Faculty of Physics, Warsaw University of Technology, ul. Koszykowa 75 00-661 Warsaw, Poland}

\date{\today}

\begin{abstract} We investigate the role of lifetime effects from resonances and emission duration tails in femtoscopy at RHIC in two Blast-Wave models.  We find the non-Gaussian components compare well with published source imaged data, but the value of $R_{out}$ obtained from Gaussian fits is not insensitive to the non-Gaussian contributions when realistic acceptance cuts are applied to models.
\end{abstract}


\maketitle


The experiments at the Relativistic Heavy Ion Collider (RHIC) have produced a series of remarkable results including the discovery of jet-quenching through the suppression of high $p_T$ particles~\cite{phnxhighpt} and the observation of hydrodynamic flow~\cite{starflow}.  The complete set of measurements point to the existence of a dense partonic medium~\cite{wp1,wp2,wp3,wp4} that evolves hydrodynamically from the time of thermalization until freeze-out.  Several hydrodynamic models have succeeded in reproducing measured spectra and flow~\cite{Huovinen2001,Kolb2003,Hirano:2002ds}, but with the exception of parameterized fits to the data~\cite{BudaLund}, they have not reproduced the space-time measurements for the particle emission region~\cite{Hirano:2002ds,Soff:2000eh,Heinz:2002un,Adler01,Adler04,Back06}.  While some the models match the systematic trends  of the longitudinal Gaussian extent, $R_{long}$, they produce an extent which is too large in the transverse component of the particle pair emission direction,$R_{out}$, too small in the orthogonal transverse direction, $R_{side}$, or both.  The ratio of these two measurement,  $R_{out}/R_{side}$, has been suggested as an indication of the duration of the freeze-out emission stage~\cite{Rischke96}, and its near unitary value for a wide range of colliding systems, geometry, and center-of-mass energy in light of the disagreement with hydrodynamic models has been called the ``HBT Puzzle''. 

The traditional spatial-temporal analysis of the particle emission region in a high energy collision pioneered by Goldhaber, Goldhaber, Lee, and Pais~\cite{gol60} assumes a parameterized source shape, usually Gaussian, which is fit to a measured enhancement in the two-particle momentum distribution at low relative momentum.  The physical basis for this technique is analagous to intesity interferemetry techniques developed by Hanbury Brown and Twiss (HBT)~\cite{hbt54} for which the technique is often named.  It has been applied mostly to pairs of identical pions emitted in both lepton and hadron collisions, but the greatest interest has been generated in the heavy ion collisions, where space-time measurements hold the promise of providing constraints on the nuclear matter equation of state.  The technique has since advanced in many ways, incorporating a variety of systematic dependencies on pair momentum, collision geometry, reaction plane and a decomposition of the source emission region into three cartesian radii, $\Rside$, $\Rout$, $\Rlong$ and cross-terms.  However, the need to parameterize the source distribution by its  ``lengths of homogeneity''~ \cite{mtscaling,heinz_99} has remained a fundamental limitation (see~\cite{annrev} for a recent review and references therein).  This limitation is underscored by the application of three-pion correlations to rule out coherent pion production as an explanation for the non-unitary value of the correlation strength, $\lambda$, which provides evidence for a complex source shape consisting of core and extended halo~\cite{Heinz:1997mr,Adams:2003vd,corehalo}.

The tools to resolve non-Gaussian components have recently been provided by a source imaging technique~\cite{imag1,imag2,imag3} and the first evidence of non-Gaussian tails have been observed by the PHENIX experiment in 200 GeV centereal Au+Au Collisions~\cite{ppg052}, thereby demonstrating the ability to image non-Gaussian sources and resolve contributions from an extended lifetime and/or resonance halo.  Understanding the role of these two contributions to an extended source distribution has important consequences for the ``HBT Puzzle" and the limitations of traditional techniques.  In this paper we will explore these limitations and their implications using two models to decouple the relative contributions from lifetime effects and resonance decays.

Our first model uses a blast-wave flow profile for the source distribution~\cite{blastwave}, but substitutes exponential time emission and Gaussian longitudinal profiles for a Gaussian in proper time and an infinite Bjorken tube.  This model also includes a resonance contribution from $\omega$ feeddown.  

The building blocks for our two-particle source are the normalized particle emission rates for the core and halo:
\begin{equation}
    D({\bf r}, t, {\bf p}) = fD_{\rm core}({\bf r}, t, {\bf p}) + (1-f)D_{\rm halo}({\bf r}, t, {\bf p})
\end{equation}
The core component (fraction = $f$), consists of a Gaussian spatial part, an exponential time profile, and a
momentum dependence arising from hydro-like Boltzman factor:
\begin{equation}
    D_{\rm core}({\bf r}, t, {\bf p}) \propto e^{-p_{\mu} u^{\mu}/T -r^2_x/2R^2_x -r^2_y/2R^2_y -r^2_z/2R^2_z -t/\tau_{fo}}
\end{equation}
Here, $f$, $T$, $R_x$, $R_y$, $R_z$, and $\tau_{fo}$ are all adjustable and are specified in the lab frame.  For this study we set $T = 165$ MeV, $f=0.5$, $R_x = R_y \equiv R_T = R_z = 4$~fm, and $\tau_{fo}=10$~fm/c. 
The flow profile is given by
\begin{equation}
    u_{\mu}(r) = \left(\cosh\eta \cosh\rho, \hat{r}_{T}\sinh\rho, \sinh\eta \cosh\rho\right)    
\end{equation}
where $\eta = \frac{1}{2}\ln\left(\frac{t+z}{t-z}\right)$ and $\rho = \rho_{0}r_{T}/R_{\rm max}$ with $\rho_{0}=0.6$.
These parameters were chosen arbitrarily since this paper is a generic study of the time profile and $\omega$ contributions to the source distribution.  For comparison, the Blast-Wave fits of Retiere and Lisa~\cite{blastwave} for central collisions converged for $R_x \approx R_y \approx 13$~fm (equivalent to a 2D rms Gaussian radius of 6.5~fm), $T \approx 110$~MeV, and $\rho_{0} \approx 0.9$. 

Following earlier work on the contribution of resonance decays to the pion source distribution~\cite{Wiedemann:1996ig,Sullivan93}, we assume that the halo is dominated by the decay of the $\omega$ resonance ($\tau_\omega = 23$ fm/c), with a fractional contribution of 50\% to the pion distribution in the region of $0.2<k_T<0.36$~GeV. Other potential candidate resonances have decay times that are much too short, e.g. the $\rho$ with lifetime $\tau_\rho = 1.3$ fm/c, or too long, e.g. the $\eta'$ with lifetime $\tau_{\eta'} = 975$ fm/c or have charged pionic decay modes with small branching fractions.  The $\omega$'s are emitted from the same core, but we allow them to propagate classically for some distance before decaying into pions:
\begin{equation}
\begin{array}{rl}
    D_{\rm halo}({\bf r}, t, {\bf p}) \propto &\displaystyle\int \dn{}{\Delta t} \dn{3}{p_\omega} P({\bf p}_\omega, {\bf
p}) e^{-\Delta t/\tau_{\omega}} \\
&\displaystyle \times D_{\rm core}({\bf r}-\frac{{\bf p}_\omega}{E_\omega}\Delta t, t-\Delta t, {\bf p_\omega}).
\end{array}
\end{equation}
We include the dominant three-body reaction $\omega\rightarrow\pi^+\pi^-\pi^0$ with a branching fraction of 88.8\% and define the probability, $P$, for finding a $\pi$ with momentum ${\bf p}$ from the decay of the $\omega$ with momentum ${\bf p_\omega}$, using standard three-body decay kinematics.   We neglect the only other charged pion decay mode of the $\omega$ which is $\omega\rightarrow\pi^+\pi^-$ with a branching fraction of 2.2\%.  This mode would only have a negligible impact on the source distribution.

This source function for the probability to emit a pair with a separation of ${\bf r}'$ in the pair CM frame is given by~\cite{imag1}
\begin{equation}
   S_{\bf P}({\bf r'})=\int \dn{}{r_0'} \int \dn{4}{R} D_1(R+r/2,{\bf P}/2) D_2(R-r/2,{\bf P}/2).
\label{eqn:sou-def}
\end{equation}

We construct the source function from this single particle source by Monte-Carlo integrating our emission function in Eq.~(\ref{eqn:sou-def}).  We work in Bertsch-Pratt coordinates \cite{pratt_90}, so that the time integral in Eq. (\ref{eqn:sou-def}) serves to move the time effects into the outward and longitudinal directions.  To compare to the 1D, angle-averaged source image for central Au-Au collisions recently measured by PHENIX~\cite{ppg052}, we keep only pairs where both pions have a pseudo-rapidity $|\eta|<0.35$ and transverse momentum $0.2$ $< k_{T} < 0.36$ GeV.  We also applied the STAR acceptance cuts in pseudo-rapidity, $|\eta|<1.0$, but found no qualitative difference.

\begin{figure}[htbp]
\begin{center}
\includegraphics[width = 0.49\textwidth, clip]{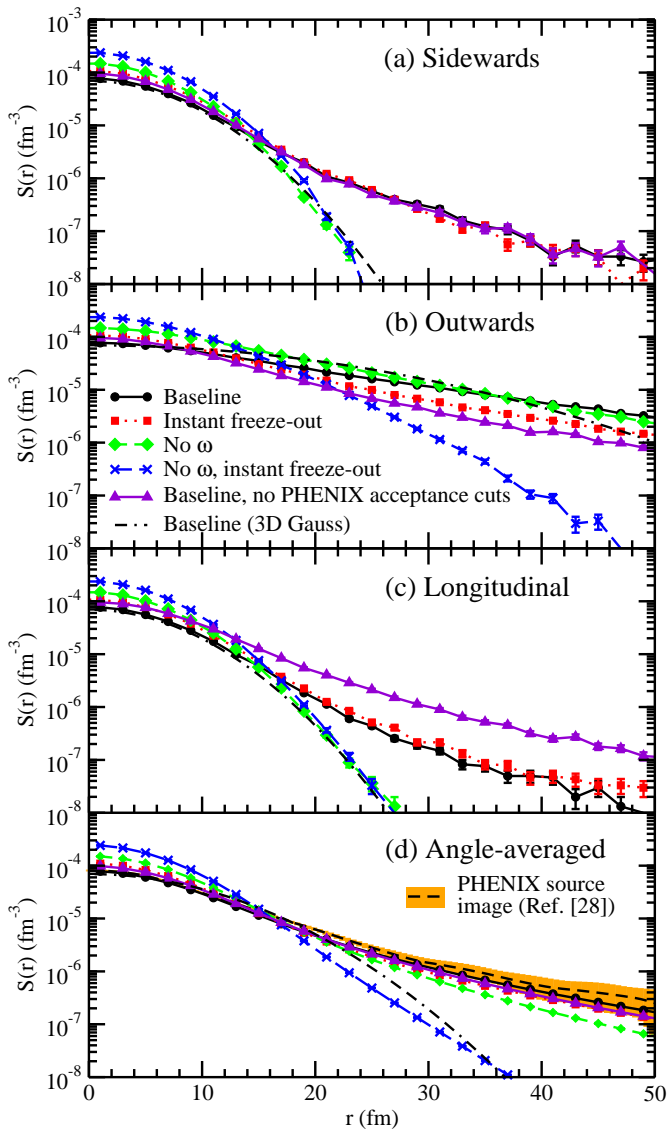}
\caption{Comparison of the sources sliced in the side, out and long directions for each of the cases we consider.  The upper panel corresponds to the sidewards direction, the middle panel to the outwards direction and the lower panel to the longitudinal direction.}
\label{compare}
\end{center}
\end{figure}

In Fig. \ref{compare}, we plot our ``baseline" model source function as 1-fm slices along the various directions in the Bertsch-Pratt coordinate system.  The top three panels show slices for $R_{side}$, $R_{out}$, and $R_{long}$, respectively, and the bottom panel shows the full angle-averged source ($\ell m = 00$ term in a spherical harmonic expansion of the source).  All distributions are normalized to one for the full distributions, and all slices are plotted in the pair center-of-mass (PCMS) frame.  Slices are also shown for various modifications to the baseline model, in which extended time distribution is turned off (``Instant freeze-out'') and/or $\omega$ contribution is removed.  The baseline model is also shown for the full $\eta$ acceptance.  The non-Gaussian shape in the sidewards direction is due entirely to the $\omega$.  The same is true for the longitudinal component, where $\eta$ acceptance cuts virtually eliminate the lifetime effects.  The outwards distribution for the $\omega$-less, instant freeze-out distribution shows the effect of kinematic boost of the PCMS frame, which varies from 1.7 to 2.8 over the transverse momentum range in this bin.  The $\omega$ and extended lifetime contribute significantly above 30~fm, although their combined contributions have a less discernible impact.  The removal of the $\eta$ acceptance cuts has the effect of reducing the impact of these contributions along the outwards slice.  The angle averaged distribution shows less sensitivity to the individual contributions to the long range source.  The angle averaged sources for the model with and without the extended lifetime effects (but with $\omega$ included) appear to be consistent with the PHENIX measurement.  

The dot-dashed line in each plot shows the 1~fm Gaussian slices for the 3D Bertsch-Pratt fit to the generated correlation function (no Coulomb) for the baseline model in the PCMS frame, and the bottom panel shows the angle-average of this function.  Fit results for all model variations are tabulated in Table~\ref{radii}.  The sidewards and longitudinal slices show a significant departure from the Gaussian shape above 15~fm, yet the outwards distribution remains Gaussian in shape out to 40~fm, significantly beyond the 30~fm that one would expect from a pure kinematic boost to the PCMS.  The sensitivity of the Gaussian fits to the outwards is also evident in the radii reported in Table~\ref{radii}.  Beginning with the pure Gaussian input source (No $\omega$, $\Delta\tau$=0) the values of $R_{side}$ and $R_{long}$  increase by $\sim$10\% when the $\omega$ is included, and $R_{long}$ increases half as much when only the emission tail is added.  In contrast, the value of $R_{out}$ increases by $\sim$60\% for either the $\omega$ or emission tail alone, and another $\sim$40\% when both are included.

\begin{table}[htdp]
\caption{Comparison of the best fit Gaussian parameters in each of the cases we consider.  We fit directly to the correlation function in each case.  In order to avoid complications due to the Coulomb correction, we generated the correlations without the Coulomb effect for the purpose of fitting.}
\begin{center}
{ \scriptsize
\begin{tabular}{r|c|c|c|c} Case              & $\lambda$ & $\Rside$ & $\Rout$ & $\Rlong$ \\
\hline \hline
No $\omega$   
$\Delta\tau=0$  & 0.998 $\pm$ 0.006 & 3.99 $\pm$ 0.02 & 5.39 $\pm$ 0.03 & 4.06 $\pm$ 0.02 \\\hline
$\Delta\tau=0$ .& 0.637 $\pm$ 0.008 & 4.32 $\pm$ 0.04 & 8.57 $\pm$ 0.10 & 4.46 $\pm$ 0.05 \\\hline
No $\omega$    & 0.954 $\pm$ 0.008 & 3.99 $\pm$ 0.03 & 8.94 $\pm$ 0.06 & 4.17 $\pm$ 0.03 \\\hline
Baseline            & 0.676 $\pm$ 0.010 & 4.36 $\pm$ 0.05 & 12.16$\pm$ 0.16 & 4.45 $\pm$ 0.05 \\\hline
Baseline 
No Cuts             & 0.719 $\pm$ 0.007 & 4.22 $\pm$ 0.03 & 6.20 $\pm$ 0.05 & 4.99 $\pm$ 0.04 \\\hline
\end{tabular}
}
\end{center}
\label{radii}
\end{table}%

\begin{figure}[htbp]
\begin{center}
\includegraphics[width = 0.49\textwidth, clip]{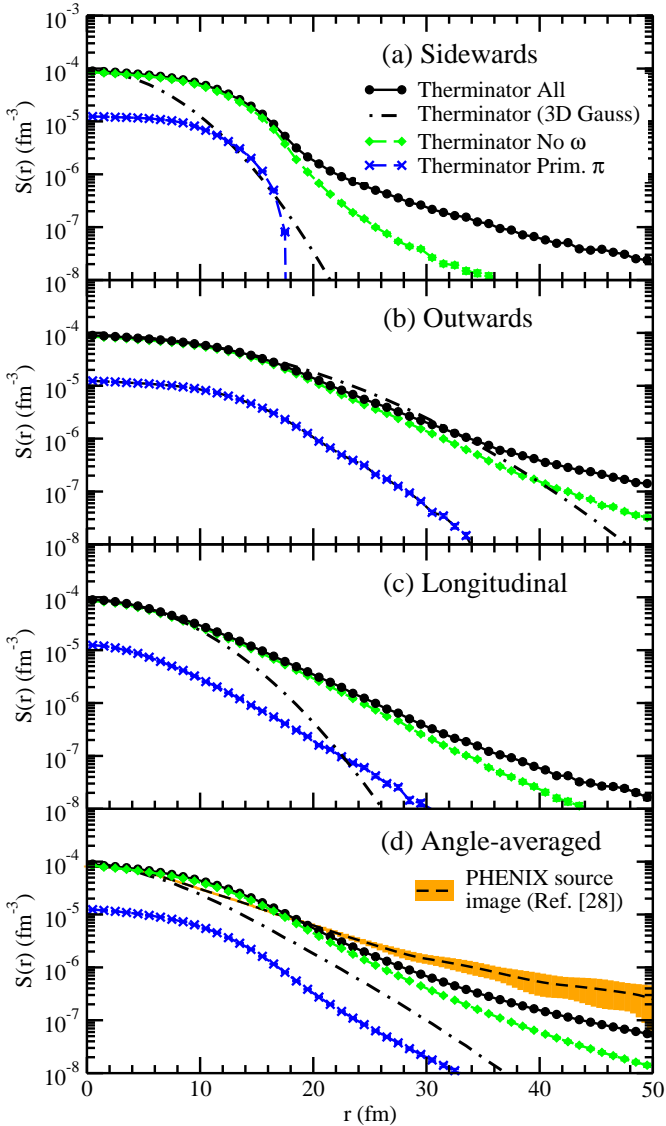}
\caption{Comparison of the Blast-Wave source from the Therminator model with and without the $\omega$ and other resonances.  We show the angle-averaged sources along with the sources sliced in the side, out and long directions.}
\label{compareBW}
\end{center}
\end{figure}

In order to investigate these effects in a more complex dynamical model with a complete resonance contribution with decay channel properties taken from the Particle Data Book~\cite{pdg} we show comparable results from the Therminator program~\cite{therminator,kisiel} in Fig.~\ref{compareBW}.  This  model includes a three-dimensional freeze-out hypersurface similar to the Blast-Wave models~\cite{blastwave} including radial flow but no emission duration.   Particles are created at this hypersurface and propagate freely (without hadronic rescattering).  The model has been shown to reproduce a large set of data in the soft sector of the RHIC collisions~\cite{therminatorFlow,therminatorStrange}, and the absolute and relative yields of particle species are taken from a thermal model which decribes RHIC data well.  The distributions shown in Fig.~\ref{compareBW} are taken from the best-fit parameters~\cite{therminator,kisiel}:  T=165.6~MeV, $\mu_b$ = 28.5~MeV, $\tau$ = 8.55 fm/c, $\rho_{\rm max}$ = 8.92~fm, $v_T$ = 0.311, using a negative slope in $r-t$ plane.  The top three panels show 1-fm wide slices along each of the Bertsch-Pratt coordinates for the same acceptance cuts, $|\eta|<0.35$ and $0.2<k_T<0.36$~GeV.  For each panel we show the full source (all resonances), the full source without the $\omega$, and the primordial pion source.  The full source is normalized to unity, and for the others we retain the relative normalization according their contribution to the two-particle source distributions.  For the Gaussian slices, we overlay the 1-fm slices from Bowler-Sinyukov fits to the full correlation source for a similar $k_T$ region (average of radii from two bins spanning $0.15<k_T<0.35$~GeV/c from~\cite{kisiel}).  To plot the outwards distributions in the PCMS frame, we multiply the LCMS $R_{out}$ fit value by 2.05, corresponding to the kinematic boost for a mean $k_T$ of 0.25~GeV/c.  Again we see that the outwards slice of the distribution shows the best agreement with a Guassian, out to $\sim$38~fm in the PCMS (or $\sim$19~fm in the unboosted LCMS), while the slices along other Bertsch-Pratt axes deviate from Gaussians around 10~fm for longitudinal and 6~fm for sidewards.  If one projects instead of slices, as was done for Fig. 11 of~\cite{kisiel}, the Gaussian shapes extend further in the sidewards and longitudinal directions.  The angle-averaged Therminator distribution including all resonances falls below the PHENIX image above 20~fm, but accounts for $\sim$30\% of the tail in this region.

We conclude from this and the previous model comparison that the $\omega$ is an essential component in the first non-Gaussian tail measured at RHIC, but there is sufficient room for contributions from non-Gaussian lifetime effects.  The observed sensitivity of the $R_{out}$ parameter to the inclusion of lifetime effects is surprising, and is contrary to our initial expectations based on an analysis without any experimental acceptance cuts in which the $R_{out}$ parameter showed little change.  Only with the application of the PHENIX/STAR $\eta$-cuts did the variation appear.  This does not rule out the presence of lifetime effects, but implies that such affects may already be partially accounted for in the standard Gaussian fits.  Indeed the fitted value of $R_{out}$ for the first model is consistent with published PCMS value for $R_{out}$ for a similar $k_T$ range~\cite{Adcox2002}.

This analysis also underscores the need to perform source imaging in three dimensions as the most promising way to disentangle the $\omega$, which has significant contributions in all directions, from the lifetime effects that are restricted to the outwards direction.  The source-imaging of kaons will provide another, more direct means to probe non-Gaussian components of primordial source lifetime.  Equally important to the task of understanding the complete space-time picture will be comprehensive set of comparisons to the full source distributions from hydrodynamic models with resonances included.

\section*{Acknowledgements}
This work was performed under the auspices of the U.S. Department of Energy by Lawrence Livermore National  Laboratory under Contract W-7405-Eng-48.  This work was supported by
the Polish Ministry of Science, grant 395/PO3/2005/29.
\nobreak

\end{document}